\documentclass[pra,superscriptaddress,onecolumn,superscriptaddress,amsmath,nofootinbib]{revtex4}

\usepackage{epsf,latexsym,bbm,bm}
\usepackage{amssymb,amsmath,amssymb}
\usepackage{times,graphics}
\usepackage{color,cancel}
\usepackage{epsfig,ulem}
\usepackage{upgreek}
\usepackage{yfonts}[1998/10/03]
\usepackage[T1]{fontenc}

\newcommand{\bra}[1]{\langle #1 |}
\newcommand{\ket}[1]{| #1 \rangle}
\newcommand{\braket}[2]{\langle #1 | #2 \rangle}
\def\6{\langle}
\def\9{\rangle}

\renewcommand{\Im}{\mathop{\rm Im}}						
\DeclareMathOperator\erf{erf}								
\DeclareMathOperator\erfc{erfc}							
	
\def\cc{{\mathcal{C}}}

\newcommand{\be}{\begin{equation}}
\newcommand{\ee}{\end{equation}}
\newcommand{\ba}{\begin{eqnarray}}
\newcommand{\ea}{\end{eqnarray}}
\newcommand{\tr}{\mathrm{tr}}

\def\bra#1{\mathinner{\langle{#1}|}}
\def\ket#1{\mathinner{|{#1}\rangle}}
\def\braket#1{\mathinner{\langle{#1}\rangle}}

{\catcode`\|=\active
  \gdef\Braket#1{\begingroup
\mathcode`\|32768\let|\BraVert\left<{#1}\right>\endgroup}}
\def\BraVert{\egroup\,\mid\,\bgroup}

\def\braket#1#2{\mathinner{\langle{#1}|{#2}\rangle}}
\def\G{{\mathcal{G}}}
\def\M{{\mathcal{G}_\S}}

\def\S{{\mathbb{S}}}

\def\H{\mathcal{H}}

\def\AB{{AB}}

\begin{document}

\title{Why should we care about quantum discord?}
\author{Aharon Brodutch}
\affiliation{Center for Quantum Information and Quantum Control, University of Toronto}
\affiliation{Institute for Optical Sciences, Department of Physics, University of Toronto}
\affiliation{The Edward S. Rogers Department of Electrical and Computer Engineering, University of Toronto}

\email{brodutch@physics.utoronto.ca}
\author{Daniel R Terno}
\affiliation{Department of Physics \& Astronomy, Macquarie University, NSW 2109, Australia}
\email{daniel.terno@mq.edu.au}

\begin{abstract}
Entanglement is a central feature of quantum theory. {Mathematical properties and physical applications of pure state entanglement make it a template to study quantum correlations. However,} an extension of entanglement  {measures to } mixed states in terms of separability   does not always correspond to all the operational {aspects}. {Quantum discord measures   allow} an alternative  way to extend the idea of quantum correlations to mixed states.  In many cases  these extensions are motivated by physical scenarios and quantum information protocols. In this chapter we discuss several settings involving correlated quantum systems, ranging from distributed gates to {detectors testing} quantum fields. In each setting we show how entanglement fails to capture the relevant  features of the correlated system, and discuss the role of discord as a possible alternative.
\end{abstract}
\maketitle

\section{Introduction}

Entanglement has been hailed as the quintessential feature of quantum mechanics. In Schr\"odinger's words it is not `` \emph{one} but rather \emph{the} characteristic trait of quantum mechanics, the one that enforces its entire departure from classical lines of thought'' \cite{Sch35}.  While its role as the only characteristic trait of quantum mechanics has been challenged, it is clear that pure bipartite entangled states play an essential role in  uniquely quantum phenomena such as Bell non-locality, steering and  teleportation \cite{horod-4,Brunner2014}. These phenomena  are not restricted to pure bipartite states, and their relation to entanglement becomes less trivial as we move to mixed states { or ensembles of pure states}. For example, it is known that not all entangled states are Bell nonlocal, steerable or useful for teleportation. Moreover,  the quantification of entanglement becomes more complicated as we step away from the pure bipartite scenario where all measures of entanglement are   functions of the spectrum of the reduced states \cite{horod-4}.  For mixed states there is a multitude of entanglement measures, matching different information processing tasks and, while all vanish on separable states, some vanish for specific entangled states. A well known example is the distillable entanglement which vanishes for bound entangled states \cite{horod-4,qinfo}.

The fact that some mixed entangled states do not {always}  exhibit properties that are directly related to the entanglement in pure states is the first hint that it does not fully capture the departure from classicality. The second hint in this direction is  that some separable mixed states  exhibit properties that are associated with entanglement for pure states.  In pure states, entanglement and classical correlation are synonymous and some properties may be mistakenly identified with {the former instead of the latter}. However, a number of phenomena  are related to correlations on the one hand,  but seem to be outside the scope of classical correlations on the other, can be observed with separable mixed states.  It has been {argued that} entanglement is only a special case of more general types of quantum correlations.  These ideas have led to a great amount of work in trying to quantify these quantum correlations using various measures that have become known under the collective name of quantum discord (see, e.g., \cite{mbc:12} and references therein)\footnote{In some cases quantum correlations and quantum discord have been used interchangeably, in other cases quantum correlations have been used as a synonym for entanglement. Here we use the term \emph{discord} when referring to discord-like quantities and \emph{quantum correlations}  when referring to a more operational aspect which may or may not relate to either discord and/or entanglement.}.

 In this chapter we present an overview of some scenarios where quantum correlations in bipartite systems are not synonymous with entanglement.   We begin with a brief discussion of entanglement in pure and mixed states, pointing out some examples where  entangled mixed states  do not have all the properties associated with  entangled pure states. We continue with a brief introduction to discord, focusing on one particular {discord measure}.  We then move on to  three examples of phenomena that involve quantum correlations and  are in some sense related to measurement disturbance.   In the first example, we examine the ability to distinguish between orthogonal pure bipartite product states.  In the second example,  we discuss more general scenarios where (the lack of) entanglement  in the input and output states fails to indicate the non-local nature of a quantum protocol. In the final  example we consider a scenario where discord is a better figure of merit than entanglement for capturing a non-classical nature of the physical system.

\section{Pure state entanglement, mixed state entanglement and discord}

\subsection{Mathematical preliminaries and notation}

We consider quantum  states that are shared between two distant parties Alice and Bob\footnote{We assume that the  identity of the subsystems is unambiguous. The extension to systems of  identical particles (where the position wave-function has to be accounted for explicitly) is mentioned in Sec.~\ref{disc}.}.  Subscripts (e.g., $A,B,\AB$) denote subsytems: for example we will consider a bipartite state $\rho_\AB$ {whose} local reduced states $\rho_A=\tr_B\rho_\AB$ and $\rho_B=\tr_B\rho_\AB$ {are controlled by} Alice and Bob, respectively. Here $\tr_K$ means a partial trace over the subsystem $K$.

 A classical probability distribution represented by the set of probabilities $\{p_k\}$ can be encoded in the quantum state $\sum_k p_k\ket{k}\bra{k}$, where $|k\9$ are the normalized orthogonal \emph{computational basis} states.  If the probability distribution is bipartite it can be encoded in the state $\sum_{k,l} p_{kl}\ket{k}\bra{k}_A\otimes \ket{k}\bra{k}_B$.

We  use entropic measures  to  quantify  most of our information theoretic quantities \cite{qinfo, wehrl:78,ct:06}. These will be based on the  von Neumann entropy, that reduces to the (classical) Shannon entropy when the states represent  classical distributions. The von Neumann entropy of a state $\rho$ is defined as $S(\rho)=-\tr\rho\log\rho$ (all logarithms used here are base $2$).  It is non-negative and vanishes only for a pure state $S(\ket{\psi}\bra\psi)=0$.  The state of maximal entropy on a  $d$ dimensional system is the   $d$ dimensional maximally mixed state $I_d$, $S(I_d)=\log d$.

An entropic measure of correlations in a quantum state $\rho$ is given by the quantum mutual information
\be
I(\rho_{AB})=S(\rho_A)+S(\rho_B)-S(\rho_\AB), \label{qmui}
 \ee
 which is one particular way to extend the corresponding classical quantity \cite{qinfo,mbc:12}. The original motivation for discord was based on the difference between various ways of extending the classical mutual information \cite{ct:06} to quantum states. A reader who is unfamiliar with the original motivation for discord is encouraged to read the original papers \cite{OllivierZurek01,HendersonVedral01} or one of the reviews on the subject \cite{mbc:12,CELERI2011}.

\textit{Fidelity} \cite{qinfo} is a measure of  {closeness} between quantum states. It is defined as $F(\rho,\sigma)=\tr\sqrt{\sqrt{\rho}\sigma\sqrt{\rho}}$ and has the following properties:
\begin{subequations}
\begin{align}
F(\ket{\psi}\bra{\psi},\ket{\phi}\bra{\phi})=|\braket{\psi}{\phi}|,\\
F(\rho_1\otimes\sigma_1,\rho_2\otimes\sigma_2)=F(\rho_1,\rho_2)F(\sigma_1\sigma_2),\label{fidprod}\\
F(\rho,\sigma)=F(U\rho U^\dagger,U\sigma U^\dagger), && \text{ For all  unitaries } U \label{fiduni} \\
F(\rho,\sigma)\le F(\Phi(\rho),\Phi(\sigma)), && \text{For all quantum channels } \Phi \label{fidnoninc}.
\end{align}
\end{subequations}
In the last equation a quantum channel $\Phi$ is represented by a completely positive trace preserving map \cite{qinfo}.

\subsection{Pure state entanglement and LOCC}

A pure bipartite state $\ket{\psi}_\AB\in\H_A\otimes\H_B$ can always be brought into the Schmidt form $\ket{\psi}_\AB=\sum_k\lambda_k\ket{\alpha_k}_A\ket{\beta_k}_B$ where $\lambda_k$ are   unique   positive  {numbers} (the Schmidt coefficients),
and $\{\ket{\alpha_k}_A\}, \; \{\ket{\beta_k}_B\}$ are complete orthogonal bases for  $\H_A$ and $\H_B$ respectively.  The state $\ket{\psi}_\AB$ is   separable (and also  a product state) if and only if it can be decomposed as $\ket{\psi}_\AB=\ket{\alpha_1}_A\ket{\beta_1}_B$, i.e it only has one non-zero Schmidt coefficient.  Pure states that are not separable are called entangled.

 The amount of entanglement in a pure state can be quantified in various ways that depend only on the Schmidt coefficients \cite{horod-4,qinfo}. Noting that $\rho_A=\sum_k{\lambda_k^2}\ket{\alpha_k}\bra{\alpha_k}$ and $\rho_B=\sum_k{\lambda_k^2}\ket{\beta_k}\bra{\beta_k}$ we see that the local states contain all the relevant information about entanglement.  Direct product states are parameterized by strictly fewer parameters than arbitrary pure  states in the same bipartite Hilbert space. Consequently, the direct product states are of measure zero in the set of all pure states.

 A pure state can be described as a state of the maximal knowledge, i.e.,  zero entropy. If a pure state is entangled, its reduced states are no longer in such a state of maximal knowledge, i.e., the local entropies are non-zero. A pure state is maximally entangled when the knowledge about the local states is minimal, i.e these states are completely mixed and {thus} have maximal entropy. In general the entanglement entropy
\be
E(\ket{\psi}\bra{\psi}_\AB)=S(\rho_A)=S(\rho_B),
\ee
is a preferred measure of a pure bipartite entanglement \cite{horod-4}. {It equals to } the Shannon entropy of the Schmidt coefficients.  Since the entropy of a pure state is zero, the mutual information is $I\left(\ket{\psi}\bra{\psi}_\AB\right)=S(\rho_A)+S(\rho_B)=2S(\rho_A)=2E(\ket{\psi}\bra{\psi}_\AB)$.

To further study the properties of pure entangled states we will describe their role in two operational tasks: Bell inequality violations and distillation. The Bell-type experiment  can be used to verify that a given state shared by Alice and Bob does not have a local realistic description in terms of hidden variables \cite{peres,Brunner2014}. A state that violates a Bell inequality is known as Bell non-local.  A  pure  state is   Bell non-local if and only if it is entangled \cite{Brunner2014}.

In the  Bell-type {experiments} Alice and Bob cannot communicate. {Scenarios where  Alice and Bob can can perform arbitrary local quantum operations on their subsystems and communicate classically, but cannot send quantum information to each other belong to the paradigm of} \emph{local operations and classical communications} (LOCC).  If Alice and Bob share some maximally entangled pairs  they can use LOCC to perform tasks that cannot be performed locally, e.g {by using} teleportation to send quantum information to each other.  Consequently, if they share an unlimited supply of maximally entangled pairs they can perform any quantum operation in a finite amount of time. If, on the other hand, they have a finite amount of partially entangled pairs, they can use LOCC to  distill them into maximally entangled pairs and  use them  for teleportation or other tasks. A  supply of entangled (but not maximally entangled) pure states can always be distilled   into a smaller supply of states that are more entangled \cite{qinfo}.

Before moving on to mixed states, we  recap a few properties of pure state bipartite entanglement that (as shown below) do not carry over to mixed states:
\begin{itemize}
\item   All separable pure states are product states (correlations $\Leftrightarrow$ entanglement).
\item Local mixed states imply a global entangled pure state (and  the local states have the same spectrum).
\item Pure product states are zero measure in the set of all pure states.
\item All pure entangled states are distillable and can be used to violate a Bell inequality.
\end{itemize}

\subsection{Mixed state entanglement}

A generic state $\rho_\AB$ (i.e a trace 1 positive-semidefinite operator on $\H_A\otimes\H_B$)  is a product state if it can be represented as $\rho_A\otimes\rho_B$. It is a separable state if it can be decomposed as
\begin{equation}
\label{eq:sep}
\rho_\AB=\sum_k\alpha_k\tau^k_A\otimes\omega^k_B
\end{equation}
(here $\{\tau^k_A\}$ and $\{\omega^k_B\}$ are sets of local  states and $\{\alpha^k\}$ is a set of probabilities).  If a state is not separable it is entangled.

{Unlike pure states,} not all separable states are product states. If a state is not a product state, it is correlated as can be verified using mutual information and the fact that $S(\rho_\AB)=S(\rho_A)+S(\rho_B) \Leftrightarrow \rho_\AB=\rho_A\otimes\rho_B$. If a mixed state is correlated it is not necessarily entangled, but if it is entangled it must be correlated. It is easy to verify whether a state is correlated or not, but it is usually difficult to to verify whether it is separable or entangled.

 The set of separable mixed states is dense.  The simplest way to see this is by showing that for small enough $p>0$ the states of the form
 \be
 \rho^{p,\psi}=p\ket{\psi}\bra{\psi}_\AB+\frac{(1-p)}{4}\openone_n
  \ee
 (where   $\openone_n$ is  the $n$ qubit identity)  are separable for any normalized  $\ket\psi_\AB$. A state of this type is called pseudo pure and is a natural state in various implementations of  quantum computing.

One interesting family of bipartite  pseudo pure states  is the family of two qubit Werner states \cite{horod-4,Werner89}.
Denote the maximally entangled singlet state $\ket{\Psi^-}=\frac{1}{\sqrt{2}}[\ket{01}-\ket{10}]$. The two-qubit Werner state is
 \begin{equation}\label{eq:Werner}
\rho^{W,p}=p\ket{\Psi^-}\bra{\Psi^-}+\frac{(1-p)}{4}\openone_2
\end{equation}
  We can think of this $\rho^{W,p}$ as a depolarized  singlet state. This state is entangled for $p>1/3$ \cite{horod-4}, but not Bell non-local for $p<0.66$  \cite{Cavalcanti2010}.

Another difference from pure states is that not all entangled mixed states can be distilled.  States that are entangled but cannot be distilled are called \emph{bound entangled}.  There are no bound entangled states for a pair of qubits or a qubit and a qutrit. A pair of qutrits {provides a simple example} using the {so-called}  tile basis and a stopper tile  \cite{Bennett99}. The tile basis   {is {formed} by the orthogonal basis states}

\begin{subequations} \label{eq:tiles}
\begin{align}
\ket{\psi_{{1}\atop{2}}}&=\tfrac{1}{\sqrt{2}}\ket{0}\otimes \big(|0\9\pm|1\9\big),\\
\ket{\psi_{{3}\atop{4}}}&=\tfrac{1}{\sqrt{2}}\ket{2}\otimes\big(|1\9\pm|2\9\big),\\
\ket{\psi_5}&=\ket{1}\otimes\ket{1},\\
\ket{\psi_{{6}\atop{7}}}&=\tfrac{1}{\sqrt{2}}\big(|0\9\pm|1\9\big)\otimes \ket{2},\\
\ket{\psi_{{8}\atop{9}}}&=\tfrac{1}{\sqrt{2}}\big(|1\9\pm|2\9\big)\otimes \ket{0},
\end{align}
\end{subequations}
and the stopper tile
\be
\ket{\psi_S}=\tfrac{1}{3}\big(\ket{0}+\ket{1}+\ket{2}]\otimes[\ket{0}+\ket{1}+\ket{2}\big).
\ee
 It is possible to show that the state
 \be
\rho_{AB}= \tfrac{1}{4}\Big(\openone_9-\!\!\sum_{i\in\{2,4,7,9,S\}}\ket{\psi_i}\bra{\psi_i}\Big),
 \ee
is bound entangled \cite{Bennett99}.

The fact that some mixed entangled states are not distillable and some cannot be used to violate a Bell inequality suggests that at least some of the properties associated with pure state entanglement are not shared by all (mixed) entangled states. In the following we discuss the opposite scenario, i.e. situations where a property that we would intuitively associate with entangled states carries over to correlated separable states.

\subsection{Discord}

The idea of quantifying quantum correlation beyond entanglement   originally {appeared in the studies of}  decoherence \cite{Zurek2000}. Within this framework it was noted that entanglement is not sufficient for capturing all quantum correlations and that some separable states retain some quantum properties.  At around the same time, a number of different versions of quantum discord  and a similar idea called the information deficit were used to quantify non-classicality in various scenarios (for a review see \cite{mbc:12}). These quantities usually vanish for one of three families of classical states, often called Quantum-Classical, Classical-Quantum and Classical-Classical (although a few vanish for more general families such as product-basis states).   A state $\rho_\AB$ is called Classical-Quantum if there is a basis on $\{\ket{a}\}$ for $\mathcal{H}_A$ and a set of states $\{\tau^a\}$ on $\mathcal{H}_B$ such that
\begin{equation}\label{eq:CQ}
\rho_\AB=\sum_a\alpha_a\ket{a}\bra{a}\otimes\tau^a.
\end{equation}
where $\alpha_a$ are probabilities that sum up to 1. The state is Quantum-Classical if it has the same structure with $A$ and $B$ swapped and it is Classical-Classical if it is both Classical-Quantum and Quantum-Classical.  These families are all measure zero in the set of all states. Various versions of  discord can be described as different ways of quantifying the `distance' from the desired family of classical states.  One way to introduce them is by calculating the difference between the quantum mutual information $I(\rho_{AB})$ that is given by Eq.~\eqref{qmui} and different measurement-dependent quantum generalizations of the classically equivalent expression \cite{ct:06,OllivierZurek01,HendersonVedral01},
\be
J^{\Pi^A}:=S(\rho_B)-S(\rho_B|\Pi^A),
\ee
{where the conditional entropy depends on the measurement on $A$ that is described by a positive operator-valued measure \cite{qinfo} $\Pi^A$ via}
\be
S(\rho_B|\Pi^A)=\sum_a p_a S(\rho_{B|a}),
\ee
{where the probability $p_a$ of the outcome $a$ is $p_a=\tr \rho_A\Pi_a$, and $\rho_{B|a}$ is the state of $B$ conditioned on obtaining the outcome $a$}.
 Different choices of optimization condition that determines the measurement selection lead to different versions of discord \cite{mbc:12},
\be
D(\rho_{AB})=I(\rho_{AB})-J(\rho_{AB}).
\ee

In this work we focus on a specific  version of discord which we call $D_3$ \cite{BTdiscord}.  It has the advantage of { being easy to calculate and  providing an upper bound on some other discord measures}. Most importantly it vanishes if and only if the states are the Classical-Quantum  states of eq. \eqref{eq:CQ}.

Given a state $\rho_\AB$ with marginals $\rho_A$ and $\rho_B$ we define the local basis $\{\ket{l_a}\}$ to be the  basis where $\rho_A$ is diagonal (note that this is not well defined when $\rho_A$ has a degenerate spectrum).  The dephasing channel $\Phi_l$ is defined as \footnote{It should be noted that this channel depends on the state, and is therefore not linear \cite{Liu2016}.}
\begin{equation}
\Phi_l(\rho_\AB)=\sum_a\ket{l_a}\bra{l_a}\rho_\AB\ket{l_a}\bra{l_a}
\end{equation}
The quantity $D_3$ is the loss of correlations under this channel
\begin{equation} \label{D3}
D_3(\rho_\AB)=I(\rho_\AB)-I\left[\Phi_l(\rho_\AB)\right]
\end{equation}

As a simple example we can consider the Werner state \eqref{eq:Werner}. If it was  classically correlated , the correlations would, in principle, be immune to decoherence, however it can easily be verified that the mutual information for a Werner state gets degraded when one of the qubits is decohered. In this sense the Werner state is always (for $p>0$) non-classically correlated. This is in-fact true for any pseudo pure state with $\ket{\psi}$ entangled.

\section{Local distinguishability and the failure of discord.}

One of the first hints that separability does not imply classicality in the context of correlations was the discovery of   \emph{non-locality without entanglement} \cite{Bennett99}. Consider a bipartite system of two qutrits and the set of nine orthonormal basis  states   of Eq. \eqref{eq:tiles}. Imagine the following task:  Alice and Bob are given one of these orthogonal states and are asked to identify which one it is, they can communicate but cannot use  any shared entanglement. Despite the fact that these are orthogonal product states the task cannot be completed deterministically. Any LOCC protocol used to identify out one of these nine states will misidentify some states with some probability. In other words,  any protocol that can perfectly identify all the nine states must include quantum communication and is in that sense non-local.  We can also say that these states are non-classically correlated although they are separable.

Now, let us assume that the apriori probability for each of the nine states is $1/9$, in such a case we can construct a density matrix $\rho_{AB}$ that represents Alice and Bob's knowledge about the unknown state. Since these states are an orthonormal basis, their equal mixture is the maximally mixed state,  $\rho_{AB}=\openone_9$. In that sense, we can see that the non-classical correlations in this scenario cannot be captured by discord in the average state since the maximally mixed state  is not  correlated \cite{BTdiscord}.   A natural approach for correcting this problem is to quantify quantum correlations in a different way for ensembles. Here we will consider a simple definition of classical ensembles which is motivated by other approaches \cite{Luo2011,Piani2014}, but requires fewer formalities.

An ensemble $\{\rho^i_{AB}\}$ is classical if and only if for any choice of non-negative coefficients $\{\alpha_i\}$, such that $\sum_i\alpha_i=1$ the state   $\rho^{\{\alpha_i\}}_{AB}=\sum_i \alpha_i\rho^i_\AB$ is classical.  It is clear that the ensemble $\{\ket{\psi_{i}}\}$ is not classical in this sense. However, neither is an ensemble that consists of two orthogonal maximally entangled pure states \cite{BTdiscord}. Now on the one hand an ensamble of two orthogonal maximally entangled states is not a classical ensemble (by the above definition), on the other hand, it is well known that any two orthogonal states can be distinguished using LOCC. Consequently, the notion of non-classically correlated ensembles  which we described above does not seem to play a role in locally distinguishing between orthogonal states.

\section{Restricted  distributed gates}

The process  of identifying an unknown  state $\ket{\phi_k}_\AB$  from the set of orthogonal states $\{\ket{\phi_i}\}$  can be described as an  isometry that takes the state $\ket{\phi_i}$ {from the space} $\H_A\otimes\H_B$  to the state $\ket{i}\otimes\ket{i}$ on a different space $\H_{A'}\otimes\H_{B'}$, where the orthogonal states $\ket{i}$ are quantum pointers to the `classical' labels. The \emph{restricted, distributed gates} paradigm \cite{BTgates,Brodutch2013} is set up along the same lines but with different restrictions.

Consider a unitary operation $\G(\rho)=U\rho U^\dagger$ (a quantum gate) and a subset of  states $\S=\{\rho^i_{AB}\}$. Now consider the family of channels $\M$ defined through
\begin{equation}
 \M(\rho^i_\AB)= \G(\rho^i_\AB), \qquad \forall \rho^i_\AB\in \S
\end{equation}
  We call such a channel  $\M$ a distributed gate if it can be implemented using LOCC.  There are situations where $\M$ cannot be distributed without shared entanglement resources,  even when both $\S$ and $\S'=\{\G(\rho^i_{AB})|\rho^i_\AB\in\S\}$ contain only separable states.  This restriction holds even when the  set $\S$  is very  small ---- in fact it can contain only two states \cite{BTgates,Brodutch2013}\footnote{Note that if $\S'$ contains only one state and this state is separable then the transformation $\M$ is trivial in LOCC.}.

We begin with the simplest case \cite{BTgates} where $\S=\{\ket{\psi_1},\ket{\psi_2}\}$ contains two non-orthogonal pure product states $\ket{\psi_i}=\ket{a_i}\ket{b_i}$, $\braket{\psi_1}{\psi_2}\ne0$. In such a case $\M$ can be implemented using LOCC if and only if $\M=\G'_A\otimes \G'_B$, where $\G'_A$ and $\G'_B$ are unitary gates. In other words if $\M$ changes correlations (classical or quantum) for any {convex combination of the two states}, $\rho_\AB=\alpha\ket{\psi_1}\bra{\psi_1}+(1-\alpha)\ket{\psi_2}\bra{\psi_2}$, then it cannot be implemented using LOCC.

The proof   of this statement is as follows. {Denote} $\ket{\psi_i^f}\bra{\psi_i^f}=\G(\ket{\psi_i}\bra{\psi_i})$. The {execution by} LOCC {of a unitary gate} implies that the output states  {are {pure and} separable},
 \be
 \ket{\psi_i^f}=\ket{a^f_i}\ket{b^f_i}, \label{purpres}
 \ee
 {as well as  the states at all intermediate steps} \cite{BTgates}.   Now, consider the protocol Alice and Bob need to use to implement the gate. Without loss of generality, we can assume that the protocol is broken into rounds where one party performs an operation and sends the classical outcomes of {their} measurement to the other party. We can also  assume  that the classical measurement results are recorded as quantum states.  Since fidelity is non-decreasing under quantum channels, Eq.~\eqref{fidnoninc}, it also cannot increase at any point   {due to unitarity of}  $\G$  which implies it {is} unchanged at the end of the process.

Assume that Alice {acts} first by {performing some operation, possibly including a measurement on her input  state that corresponds to a classical message $k$ that she sends to Bob.  When averaged over many implementations of the protocol, it results in a}  channel $\Phi_A$. As a result of  Eq.~\eqref{fidprod} the fidelity
\be
F\big(|a_1\9\6{a_1}|,|a_2\9\6{a_2}|\big)=|\6a_1|a_2\9|=F\big(\Phi_A(\ket{a_1}\bra{a_1}),\Phi_A(\ket{a_2}\bra{a_2})\big),
 \ee
 {is preserved}.  Alice's state is now {is either of}
 \begin{subequations}
 \begin{align}
 \Phi_A(\ket{a_1}\bra{a_1})=\sum_kp_1^k|a_1^k\9\6a_1^k|\otimes\ket{k}\!\bra{k}^{A'},\\  \Phi_A(\ket{a_2}\bra{a_2})=\sum_kp_2^k|a_1^k\9\6a_1^k|\otimes\ket{k}\!\bra{k}^{A'},
 \end{align}
 \end{subequations}
  where { $p_i^k$ are the probabilities of obtaining the outcome $k$ given the state $i=1,2$,  and the subsystem} $A'$ holds the classical information to be sent to Bob. Since the pointer states on $A'$ are orthogonal,  {the fidelity satisfies}
 \be\label{eq:fidblock}
 |\6a_1|a_2\9|=F\big(\Phi_A(\ket{a_1}\bra{a_1}),\Phi_A(\ket{a_2}\bra{a_2})\big)=\sum_k\sqrt{p_1^kp_2^k}\,\big|\6a_1^k|a_2^k\9\big|,
 \ee
 {However, since
 \be
 \sum_k\sqrt{p_1^k p_2^k}\leq 1,
 \ee
  Eq.~\eqref{eq:fidblock} cannot be satisfied   unless  either {the probability distributions coincide} (which implies that Alice has no relevant information to send Bob\footnote{If Alice has no relevant information to send Bob, then by symmetry Bob cannot have any relevant information to send Alice, and the protocol should not involve any communication.}) { and
 \be
 |\6a_1^k|a_2^k\9\big|=|\6a_1|a_2\9|,\qquad \forall k,
 \ee 
or  there must be some $l$ with   }
\be
|\6a_1^l|a_2^l\9\big|>|\6a_1|a_2\9|.
\ee
However, if Alice gets the result $l$ (that she will send to Bob) {and then the two parties proceed with the successful implementation of the protocol}, {they must deterministically decrease the fidelity in at least one stage on the way} to the final state. {This contradicts the non-decreasing of fidelity in quantum channels,} Eq.~\eqref{fidnoninc}. The conclusion is that Alice gets no useful information during the measurement and has nothing to send Bob. Consequently the overall  {transformation} must be {implemented by} local unitary operations.

In the general case,  it can be shown \cite{Brodutch2013} that if $\S$ contains only two states: $\rho$ and the maximally mixed state, then an LOCC $\M$ cannot change the correlations in $\rho$ unless there is some measurement that leaves $\rho$ invariant.  {This suggests that the maximally mixed state {may} play an important role in increasing the quantum resources required by a quantum protocol.}

\section{Discord and Unruh-DeWitt detectors}

Various discord-like quantities were calculated in a number of problems of relativistic quantum information \cite{rqi,mbc:12}. The scenario we consider below is interesting from several points of view. The state $\rho_{AB}$ of the two detectors that are used to characterize the vacuum entanglement belongs to the family of  the X-states at all orders of the perturbation theory; the discord $D_3$ is a natural  quantity to characterize quantumness of correlations; correlations and discord persist in the region of  strictly zero entanglement.

\subsection{The model}
\label{The detector model}

From the point of view of local observers the vacuum state of any quantum  field is entangled, and thus localized vacuum fluctuations are correlated \cite{rqi}. It was demonstrated that vacuum correlations  measured by local inertial observers can, in principle,   violate Bell-type inequalities \cite{Summers:1985}. Further, it is known that localized particle detectors can extract entanglement form the vacuum state of a quantum field, even while remaining spacelike separated \cite{valen:91,reznik:03,hll:12}.

An Unruh-DeWitt detector is a two-level quantum systems that interacts with (a   real massless) scalar field $\phi$ via a monopole coupling \cite{Crispino:2008}.  It is a popular tool in analysis of entanglement in quantum fields. The time-dependent interaction Hamiltonian in the interaction picture is given by
\begin{align}
H_I \left(\tau\right) = \lambda\! \left(\tau\right) \left(e^{i\Omega \tau}\sigma^+ + e^{-i\Omega\tau} \sigma^- \right)  \phi\left[x(\tau)\right], \label{InteractionHamiltonian}
\end{align}
where $\tau$ is the proper time of the detector, $\lambda\left(\tau\right)$ is a weak time-dependent {coupling parameter} that controls the strength and length of the interaction, $\Omega$ is the energy gap between the ground state $\ket{0}_d$ of the detector and its excited state $\ket{1}_d$, $\sigma^{\pm}$ are SU(2) ladder operators that act on the state of the detector according to $\sigma^+ \ket{0}_d = \ket{1}_d$, $\sigma^- \ket{1}_d = \ket{0}_d$, ${(\sigma^{\pm})}^2=0$, and $\phi\left(x(\tau)\right)$ is the field evaluated along the trajectory of the detector.

It is convenient to parameterize the time evolution by the common coordinate time $t$ \cite{mmst:16}. We {express the coupling parameter as $\lambda(t) =  \epsilon_0 \epsilon(t)$, where $\epsilon_0\ll 1$ is the coupling strength and $\epsilon (t) = e^{-t^2/2\sigma^2}$ is a Gaussian switching function.}

Prior to the interaction the detectors had been  in their ground states $\ket{0}_A$ and $\ket{0}_B$, and the field  in the vacuum state $\ket{0}$, hence the initial joint state of the two detectors and field was given by $\ket{\Psi} = \ket{0}_A\ket{0}_B\ket{0}$. The unitary evolution of the detectors-field system is given by
\begin{align}
U = \hat{T} e^{-i\int\!dt  [ H_A(t) +H_B(t)] }   , \label{jointUnitary}
\end{align}
where $\hat{T}$ denotes time ordering and the Hamiltonians $H_A$ and $H_B$ (that are given by Eq.~\eqref{InteractionHamiltonian} each) describe the field interaction with detectors $A$ and $B$, respectively.

 The joint state of the two detectors  is $\rho_{AB} = \tr_\phi [ U \ket{\Psi} \!\bra{\Psi} U^\dagger]$, where the trace is over the {field degrees of freedom}.
 It is possible to show that \cite{mmst:16} at all orders of perturbation theory the density matrix
has the form of an  X-state \cite{ali:10}
\begin{align}
 \rho_{AB}=\begin{pmatrix}
r_{11} & 0 & 0 & r_{14}e^{-i\xi} \\
0 &r_{22} & r_{23}e^{-i\zeta} & 0 \\
0 & r_{23}e^{i\zeta} &r_{22} & 0\\
r_{14}e^{i\xi} & 0 &0 & r_{44},
\end{pmatrix}, \label{rhoAB}
\end{align}
in the basis $\left\{ \ket{00},  \ket{01}, \ket{10}, \ket{11} \right\}$ where $\ket{ij} = |i\9_A |j\9_B$, and all the coefficients $r_{ij}$ are positive.  Since $\rho_{AB}$ is a valid density matrix, the normalization condition $\sum_i r_{ii}=1$, and the following two positivity conditions must be satisfied:
\begin{align}
r_{11}r_{44}\geq r_{14}^2, \qquad r_{22}r_{33}\geq r_{23}^2.
\end{align}

A useful {parametrization} of this matrix that explicitly separates the local and nonlocal quantities is
\begin{align}
\rho_{AB}=\begin{pmatrix}
1- A-B+E & 0 & 0 & X \\
0 &B-E & C & 0 \\
0 & C^* &A-E & 0\\
X^* & 0 &0 & E
\end{pmatrix}, \label{TwoDetectors}
\end{align}
{where $A$ and $B$ are the probabilities that either detector $A$ or $B$ are excited after the interaction with the field, and the other parameters are functions of the properties of both detectors.} Indeed, tracing out either of the detectors in the state $\rho_{AB}$, say detector $B$, results in the state $\rho_A$ of detector $A$
\begin{align}
\rho_{A}=
\begin{pmatrix}
1- A& 0  \\
0 & A
\end{pmatrix}, \label{SingleDetector}
\end{align}
in the basis $\{ \ket{0}_A, \ket{1}_A \}$.

 To simplify the exposition we consider  the case of two  identical detectors, i.e., $H_A=H_B$, at rest at the distance $L$ from each other   in the  Minkowski spacetime. We find
 \cite{mmst:16} the matrix elements of $\rho_{AB}$ to be
 \begin{align}
A &= \frac{\epsilon_0^2}{4 \pi}\left[e^{-\sigma^2 \Omega^2} -  \sqrt{\pi} \sigma \Omega \erfc \left(\sigma \Omega\right)\right]+ \mathcal{O}\!\left(\epsilon_0^4\right), \label{AinMinkowski} \\
X &=  \frac{\epsilon_0^2}{4\sqrt{\pi}} \frac{\sigma}{L} i  e^{-\sigma ^2 \Omega ^2 - \frac{L^2}{4\sigma^2}}   \left[1 + \erf\left(i\frac{L}{2\sigma}\right)\right]+ \mathcal{O} \! \left(\epsilon_0^4\right), \label{XinMinkowski}\\
C &=  \frac{\epsilon_0^2 }{ 4 \sqrt{\pi}} \frac{\sigma}{L} e^{-\frac{L^2}{4\sigma^2}} \Bigg(    \Im \left[ e^{i  \Omega L} \erf\left( i \frac{L}{2\sigma} + \sigma \Omega\right) \right]  -    \sin\left( \Omega L\right)  \Bigg) + \mathcal{O}\!\left(\epsilon_0^4\right) \label{Cexp},\\
E  &= |X|^2 + A^2 + 2 C^2 + \mathcal{O} \! \left(\epsilon_0^6\right), \label{Eexp}
\end{align}
where $\erf(z)$ is the error function, $\erfc(z) = 1 - \erf(z)$. When the  distance $L$ between  detectors increases the total state approaches the direct product of the density matrices of the individual detectors,
{that is}
\be
X\to 0, \qquad C\to 0, \qquad E\to A^2. \label{sepAB}
\ee

\subsection{Information-theoretical properties of the joint state}
\label{Informational properties of the joint state}

 Application of the Peres--Horodecki criterion \cite{horod-4,qinfo} shows that the X-states are entangled if and only if either of the alternatives
\begin{align}
r_{14}^2>r_{22}r_{23}, \qquad r_{23}^2>r_{11}r_{44},
\end{align}
holds.

\begin{figure}[htbp]
\includegraphics[width=0.5\textwidth]{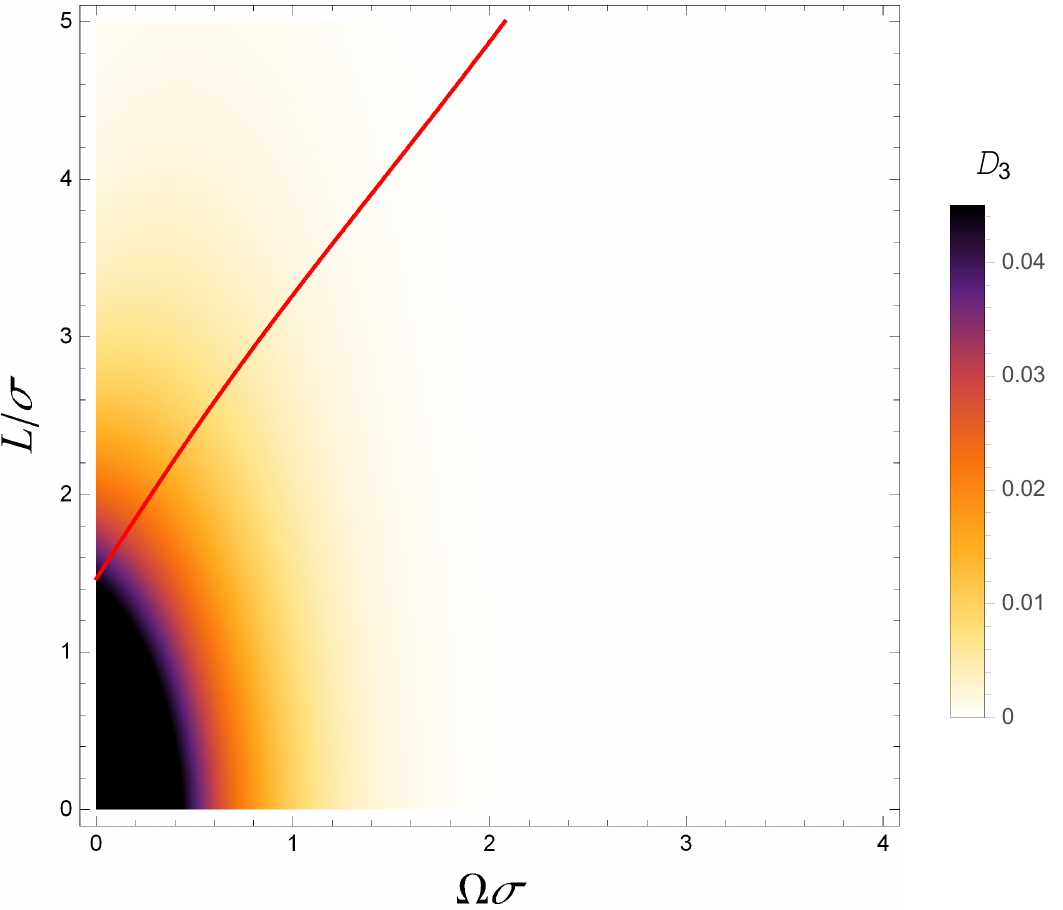}
\caption{The discord $D(\rho_{AB})/\epsilon_0^2$ as given by Eq.~\eqref{d3}. The domain of zero entanglement lies above the red line, {but there is no reason to suspect any qualitative change in the physics in the vicinity of this line. } }
\label{setup}
\end{figure}

For two identical detectors in the state $\rho_{AB}$ given in Eq.~\eqref{TwoDetectors}, these conditions are equivalent to
\begin{align}
|X|-A+\mathcal{O}(\epsilon_0^4)>0,\qquad |C|-\sqrt{E} +\mathcal{O}\!\left(\epsilon_0^4\right)  >0, \label{pptcond}
\end{align}
respectively. However, Eqs.~\eqref{Cexp} and \eqref{Eexp} ensure that the second condition is never satisfied. The concurrence \cite{horod-4,qinfo}
\begin{align}
\cc=2\,\mathrm{max}\left(0,|X|-A+\mathcal{O}\!\left(\epsilon_0^4\right) \right), \label{concurrenceMinkowski}
\end{align}
is non-zero if and only if $r_{14}>r_{22}$.  This is the area below the red line on Fig.~1.

{Concurrence and other entanglement measures are inaccessible by local measurements.   Instead we  focus on the correlation between the detectors $A$ and $B$. We characterize the {measurement}  results by random variables $r_A$ and $r_B$, respectively, with $r_A, \ r_B \in \{0,1\}$.  The correlation between these variables is given by
\be
\mathsf{corr}_{AB} =\frac{\mathsf{cov}_{AB}}{\sigma_A\sigma_B} = \frac{E-AB}{\sqrt{A(1-A) B(1-B)}}=\frac{|X|^2+2C ^2}{A}+\mathcal{O}\!\left(\epsilon_0^4\right), \label{corrGeneral}
\ee
where $\mathsf{cov}_{AB} :=\6 r_A r_B\9-\6{r_A}\9\6{r_B}\9$ is the covariance between $r_A$ and $r_B$ and $\sigma_A^2 = \mathsf{cov}_{AA} $ and $\sigma_B^2 = \mathsf{cov}_{BB}$ are the variances associated with $r_A$ and $r_B$.

Eq.~\eqref{pptcond} implies that the state $\rho_{AB}$ is not entangled when the local {terms dominate the bi-local effects},   i.e. $A>|X|$. However,  {the state  still contains} non-classical correlations that are characterized by quantum discord.

The measurement on the Unruh-DeWitt detector precisely selects the eigenbasis of Eq.~\eqref{SingleDetector}, making the discord $D_3$ the preferred measure.  The measurement of $A$ is a standard projective measurement in the eigenbasis of the reduced state $\rho_A$, and
\begin{align}
D_3(\rho_{AB})=S(\rho_{AB}^*)-S(\rho_{AB}),
\end{align}
where $S(\rho)$ is the von Neumann entropy of the state $\rho$ and
\be
\rho_{AB}^*=\sum_a\rho_{B|a}\otimes |a\9\6 a|,
\ee
i.e., $\rho_{AB}^*$ is the average state of the joint system after the eigenbasis projective measurement on $A$ \cite{BTdiscord}.

The discord $D_3$ stands out {for the same reason that the entanglement measures are unobservable}. The optimization procedures required by other measures are unavailable: no basis other than the standard $\left\{\ket{0}, \ket{1}\right\}$  is accessible.   A straightforward calculation shows that
\be
D_3(\rho_{AB}) =\big(A+C\big)\log\big(A+C) +\big(A-C\big)\log\big(A-C\big) -2 A \log A +\mathcal{O}(\epsilon_0^4),\label{d3}
\ee
hence quantum correlations persist for any finite separation of the detectors.

\section{Discussion}\label{disc}

While entanglement is a central feature of quantum mechanics {and a fundamental resource in quantum information processing, it is not always directly related to interesting qualitative features of correlated quantum systems, be it  open quantum systems correlated with the environment or quantum phase transitions in correlated many body systems. Extensions of the quantitative measures of pure state entanglement into mixed states, while very useful in many scenarios\footnote{Mixed state entanglement monotones are a particularly good quantities  in scenarios where entanglement is consumed as a resource.}, do not capture all the richness  of non-classical correlations. These are the settings were the discord-like quantities find their use}.

The three examples described are only a small sample of the vast work regarding the role of correlations (quantum or classical) in various non-classical scenarios (see \cite{mbc:12} for a review). In many cases the role of correlations is still being explored, and in others the relation seems to have failed (see for example the work on correlations and complete positivity \cite{arXiv:1012.1402, Brodutch2013a,Buscemi}).

 {Another area that generated a burst of interest with respect to quantum correlations is quantum computing and the difficulty of simulating  large quantum systems.  It is known that in general it is possible to efficiently simulate the dyanamics of a many-body system when the state is pure and  the entanglement (as measured by the Schmidt rank over all bipartitions)  scales at most logarithmically with the system size at all times \cite{sim-cq}. Similarly it is known that a quantum circuit can be efficiently simulated (classically) when the correlations between the qubits are restricted to blocks of a constant size \cite{JL}. Both of these results imply that pure state dynamics should be easy to simulate when the systems are separable. It has, however been speculated \cite{JL} that this is not true for mixed states. In general there is growing evidence \cite{DV,DSC,BBM} that mixed state dynamics may be difficult to simulate even when the systems are separable over most bi-partitions. }

{One particularly unexplored, but potentially important area, is the study of quantum correlations in the system of identical particles. Entanglement of identical particles, particularly in many-body systems \cite{ent-id0}, has features distinct from that of the identical particles, and poses more open questions. Discord-like quantities are also much less understood beyond few-fermion systems \cite{majtey:13}. On the other hand, quantum simulations of many-body system \cite{sim-q}, may resolve the problem of the exponential scaling of resources needed to calculate, e.g., energies of atoms or molecules, with their size \cite{fulde:16}.  {While correlations seem to be at the root of the requirement for exponential resources, it is unclear how to best quantify the correlations in order to provide the best figure of merit for the difficulty of simulating the many-body system. } This still remains to be investigated, particularly in light of the classical cumulant-based methods of  approximating many-electron wave functions that raise the possibility of breaking this exponential wall \cite{fulde:16}.

It is clear that ideas regarding pure state entanglement do not always carry forward to  mixed state entanglement and that in many cases a more general (and sometimes more restrictive) class of states must be considered.  The question we should be asking is therefore not \emph{why discord?} but rather \emph{when discord?}. We hope that this brief review further stimulates work in that direction.

\begin{acknowledgments}
A part of this work was done when AB was at the Institute for Quantum Computing and the Department of Physics and Astronomy at the University of Waterloo. AB was supported by NSERC, Industry Canada, CIFAR and a fellowship from the Center for Quantum Information and Quantum Control.
\end{acknowledgments}

\end{document}